\documentclass[conference]{IEEEtran}
\IEEEoverridecommandlockouts
\usepackage{cite}
\usepackage{amsmath,amssymb,amsfonts}
\usepackage{algorithmic}
\usepackage{graphicx}
\usepackage{multirow}
\usepackage{textcomp}
\usepackage{xcolor}
\usepackage{hyperref}
\ifCLASSOPTIONcompsoc \usepackage[caption=false,font=normalsize,labelfon
t=sf,textfont=sf]{subfig}
\else
\usepackage[caption=false,font=footnotesize]{subfig}
\fi
\def\BibTeX{{\rm B\kern-.05em{\sc i\kern-.025em b}\kern-.08em
    T\kern-.1667em\lower.7ex\hbox{E}\kern-.125emX}}
    
\graphicspath{{./}}
    
\begin{document}

\title{ScaDLES: Scalable Deep Learning over Streaming data at the Edge}

\author{\IEEEauthorblockN{Sahil Tyagi}
\IEEEauthorblockA{\textit{Department of Intelligent Systems Engineering} \\
\textit{Luddy School of Informatics, Computing and Engineering} \\
\textit{Indiana University Bloomington} \\
Indiana, USA \\
styagi@iu.edu}
\and
\IEEEauthorblockN{Martin Swany}
\IEEEauthorblockA{\textit{Department of Intelligent Systems Engineering} \\
\textit{Luddy School of Informatics, Computing and Engineering} \\
\textit{Indiana University Bloomington} \\
Indiana, USA \\
swany@iu.edu}
}

\maketitle

\begin{abstract}
	Distributed deep learning (DDL) training systems are designed for cloud and data-center environments that assumes homogeneous compute resources, high network bandwidth, sufficient memory and storage, as well as independent and identically distributed (IID) data across all nodes.
However, these assumptions don't necessarily apply on the edge, especially when training neural networks on streaming data in an online manner.
Computing on the edge suffers from both systems and statistical heterogeneity.
Systems heterogeneity is attributed to differences in compute resources and bandwidth specific to each device, while statistical heterogeneity comes from unbalanced and skewed data on the edge.
Different streaming-rates among devices can be another source of heterogeneity when dealing with streaming data.
If the streaming rate is lower than training batch-size, device needs to wait until enough samples have streamed in before performing a single iteration of stochastic gradient descent (SGD).
Thus, low-volume streams act like stragglers slowing down devices with high-volume streams in synchronous training.
On the other hand, data can accumulate quickly in the buffer if the streaming rate is too high and the devices can't train at line-rate.
In this paper, we introduce \emph{ScaDLES} to efficiently train on streaming data at the edge in an online fashion, while also addressing the challenges of limited bandwidth and training with non-IID data.
We empirically show that \emph{ScaDLES} converges up to $\mathbf{3.29\times}$ faster compared to conventional distributed SGD.
\end{abstract}



\section{Introduction}

With the advent of big data and IoT, the number of smart devices has grown exponentially over the years.
These devices capture data across a wide range of modalities, such as image/video in smartphones and surveillance camera feeds, audio and speech from smart speakers, text/language on phone/tablet keyboards etc.
The data collected on the devices can either be moved to a centralized server in the cloud or persist locally.
Local storage is practical when network bandwidth is limited and data privacy is a concern.

Locally storing data presents its own challenges due to limited capacity on edge/fog devices.
The problem is exacerbated on devices with high-inflow streaming data.
The data lifetime is also influenced by device streaming rates as high-volume streams may require more frequent storage purge or handling via other means.
Commercial solutions offer data storage in the cloud for finite time, but this violates data privacy, incurs high communication cost of data movement and the subscription-based cost to store that data.
\emph{Distributed deep learning (DDL)} typically assumes centralized data, where each process/device samples training data in an IID fashion at every iteration.
However, this is not necessarily true for streaming data which can be skewed not just in volume, but can be unbalanced and have non-IID distribution as well.
Another consequence of training on devices with varying flow-rates is that high-inflow devices may have to wait on low-inflow ones until they gather enough samples corresponding to the mini-batch set for training.
Thus, devices with low-volumes of streaming data can be essentially perceived as \emph{stragglers} that slow down distributed training.

In DDL, gradients computed locally are aggregated into a global update which is propagated back to the devices before proceeding to the next iteration. 
The size of the gradients communicated is of the same scale as the number of trainable parameters in the network, which can span over hundreds of millions or even billions for modern language and vision models.
Using single-precision ($32$-bit) floats to represent gradients means that hundreds of megabytes or even gigabytes of data needs to be exchanged at every iteration.
Thus, heterogeneity in data inflow among devices, unbalanced-ness in device-local data, finite memory/storage and limited bandwidth violate assumptions of conventional distributed training designed for HPC and cloud.

In this paper, we build a streaming-based distributed training framework cognizant of the aforementioned issues that we call \emph{\textbf{ScaDLES:}} \textbf{\{Sca\}}lable \textbf{\{D\}}eep \textbf{\{L\}}earning over \textbf{\{S\}}treaming data at the \textbf{\{E\}}dge.
\emph{ScaDLES}\footnote{Code available at \href{https://github.com/sahiltyagi4/ScaDLES}{https://github.com/sahiltyagi4/ScaDLES}} is designed to train across devices with heterogeneous volumes of streaming data in an online manner.
Instead of waiting for all workers to accumulate enough samples corresponding to the mini-batch, we choose a variable min-batch size for each device based on its streaming rate.
As a result, there is no additional wait-time on account of low-volume devices.
To aggregate gradients across workers, we perform \textit{weighted aggregation} such that a device with a larger batch size is weighted more than those with smaller batches.
We empirically show how this weighted gradient aggregation approach converges faster than typical distributed SGD.

To tackle the issues of limited memory, storage and expensive disk IO, we compare two simple data storage policies: \textbf{Stream Persistence} and \textbf{Truncation}.
We simulate streams and implement these policies with Apache Kafka \cite{b1}, a popular distributed stream processing platform.

Lastly, to deal with limited bandwidth and high communication cost of gradient reduction, we propose an adaptive compression technique where we scale the compression ratio based on gradient variance and adjusting to critical regions \cite{b4} in the training phase.
We apply this adaptive method on Top-\textit{k} gradient sparsification \cite{b5}.
\emph{ScaDLES} works in an online, black-box manner that we validate by simulating streams with different degrees of heterogeneity, both on IID and non-IID data, and compare performance with conventional distributed SGD.

\section{Challenges in Streaming DL}\label{sec:challenges}

DDL training on streaming data presents unique challenges that can severely impact training time and/or convergence quality.
Using data streams simulated on Kafka and Pytorch's \cite{b6} distributed data-parallel \cite{b7} module, we observe the effects of heterogeneous streams, skewness in training data, limited memory/storage and communication cost of synchronizing model updates on the overall training time and model convergence.

\subsection{Heterogeneity in device streaming rates}\label{subsec:heterostream}

In conventional DDL, multiple devices train a local model replica on partitions sampled from the entire training dataset, and aggregate gradient updates at the end of each iteration either via parameter servers \cite{b8} or Allreduce using communication libraries like Open MPI \cite{b9} and NCCL \cite{b10}.
With distributed SGD, parameter update $\mathit{w}$ at iteration $(\mathit{t}+1)$ for $\mathit{N}$ devices optimizing loss function $\mathit{\mathcal{L}(\cdot)}$ on a sample $\mathit{x_{i}}$ of size $\mathit{b_{i}}$ from distribution $\mathcal{X}_{i}$ and learning rate $\mathit{\eta}$ is given by Eqn. (\ref{eqn:distsgd}).


\begin{equation}
w_{t+1} = w_{t} - \eta \dfrac{1}{N} \sum_{n=1}^{n=N}{\dfrac{1}{|b_{i}|} \sum_{i \in b_{i}} \dfrac{\partial}{\partial w_{t}} \mathcal{L}(x_{(i,n)},w_{t})}
\label{eqn:distsgd}
\end{equation}

Each device trains on the same mini-batch size $\mathit{b}$, making global batch-size $\mathit{N \cdot b}$.
However, when dealing with streaming data,  devices can have different streaming rates.
Devices with high-volume streams can readily collect $\mathit{b}$ samples, while those with sparse inflow rates need to wait until samples equal to $\mathit{b}$ are collected.
With an inflow rate of $\mathit{p}$ samples/sec., a device would have to wait about $\mathit{(b/p)}$ seconds before proceeding to perform forward-backward pass.
We consider such variances in streaming rates among devices as \emph{\textbf{streaming heterogeneity}}.
Heterogeneity can be inter or intra-device as well; the streaming rate on a device itself can vary based on traffic, usage, time of day, etc.
To understand how streaming heterogeneity can affect wall-clock time due to latency incurred while gathering a mini-batch, we sample streaming rates from different distributions and compute the latency incurred to collect different batch sizes.

\begin{table}[!t]
\renewcommand{\arraystretch}{1.3}
\caption{Devices in a cluster sampled with varying streaming rates}
	\centering
	\begin{tabular}{|c|c|c|c|}
	\hline
	\bfseries Distribution & \bfseries Sample set & \bfseries Mean & \bfseries Std. Dev. \\
	\hline
	\multirow{2}{*}{Uniform} & $\mathit{S_{1}}$ & 38 & 24 \\
	\cline{2-4}
	& $\mathit{S_{2}}$ & 300 & 112 \\
	\hline
	\multirow{2}{*}{Normal} & $\mathit{S_{1}'}$ & 64 & 24 \\
	\cline{2-4}
	& $\mathit{S_{2}'}$ & 256 & 28 \\
	\hline
	\end{tabular}
\label{distribution_table}
\end{table}

In Table \ref{distribution_table}, we use two sets each of uniform and normal distribution to sample streaming rates for devices.
These two distributions capture inflow heterogeneity that we typically expect to see in real-world settings.
Uniform distribution samples evenly across a given range, thus giving more heterogeneous streaming rates.
On the other hand, rates sampled in normal distribution are centered around the mean so it resembles a more homogeneous setting w.r.t the device streaming rates.
Sets $\mathit{S_{1}}$ and $\mathit{S_{1}'}$ have a smaller mean as well as variance, while $\mathit{S_{2}}$ and $\mathit{S_{2}'}$ represent a higher mean and larger standard deviation.
$[\mathit{S_{2}}, \mathit{S_{2}'}]$ denote higher streaming rates compared to $[\mathit{S_{1}}, \mathit{S_{1}'}]$.

\begin{figure}
\subfloat[Latency in $\mathit{S_{1}}$ and $\mathit{S_{2}}$]{\includegraphics[width=0.25\textwidth]{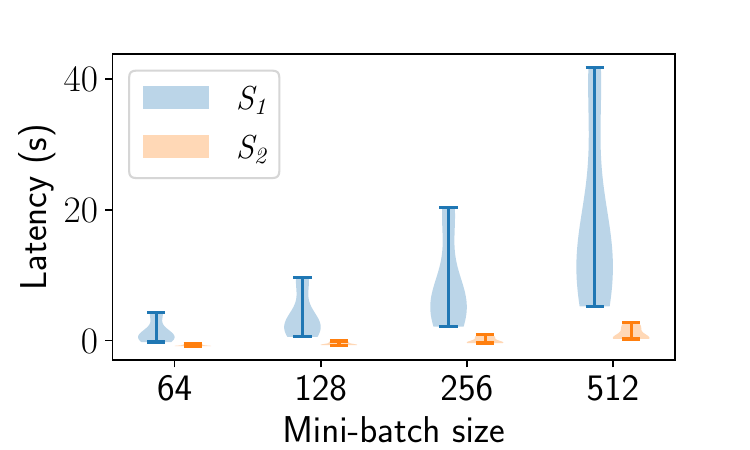}
\label{uniformdist}}
\subfloat[Latency in $\mathit{S_{1}'}$ and $\mathit{S_{2}'}$]{\includegraphics[width=0.25\textwidth]{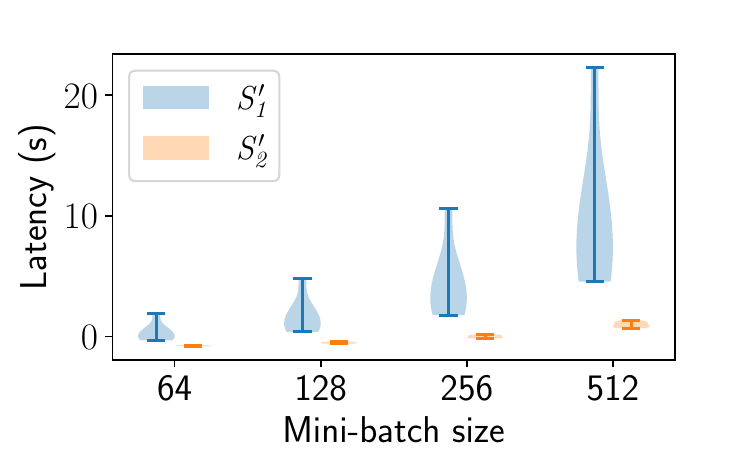}
\label{normaldist}}
\caption{Streaming latency across batches when device stream-rates are sampled from different distributions.} \label{heterogenstreams}
\end{figure}

The batch size is an important \emph{hyperparameter} in deep learning, i.e., a factor that influences convergence in neural networks.
A small batch size cannot be efficiently parallelized, while a very large batch size increases generalization error \cite{b11}.
For now, we don't take these considerations into account and only see the latency incurred to gather different batch sizes when we sample streaming rates from the described distributions.
Fig. \ref{heterogenstreams} shows the streaming latency across each set for different batches.
Latency increases with larger batches as more training samples need to be collected.
\textit{Thus, the device with the lowest streaming rate (and maximum latency) effectively becomes a straggler in synchronous training as other devices wait on it to gather a mini-batch, perform computation and send its local gradients for reduction.}

\subsection{Data skewness in deep learning}

While collaboratively training models, data on a device can be skewed either in volume, properties, or both.
For unbalancedness due to volume, imagine a traffic surveillance system where devices capture identically distributed data like frames of individual vehicles (car, bike, trucks, etc.), but the volume of data on each device varies with the traffic density on the route where the camera is installed.
Skewness due to data properties is introduced when the distribution of device-local data varies significantly from the overall data distribution. 
For example,  a vehicle recognition model running on a video surveillance system installed in a subway captures images of trains, while devices installed on the airport cover flying vehicles only.
Thus, training data has non-IID distribution as it has partial labels only (like a train or a plane).
Privacy-sensitivity and large volumes of data on constrained networks make it unfeasible to move it to a centralized location like the cloud.

\begin{figure}
\centerline{\includegraphics[width=0.3\textwidth]{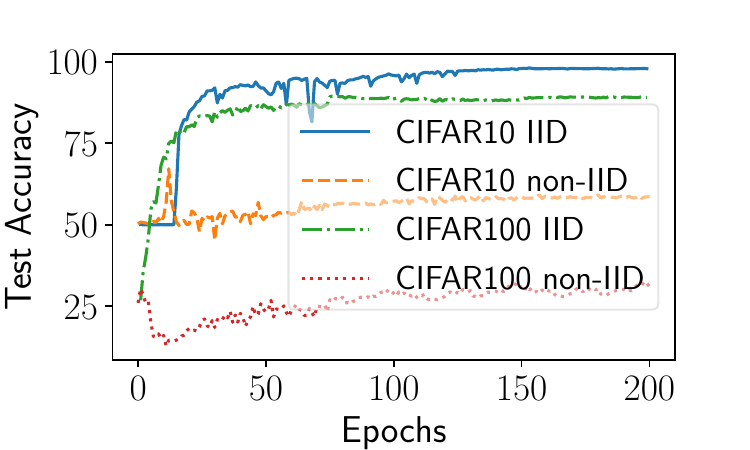}}
\caption{Test accuracy for ResNet152 on CIFAR10 and VGG19 on CIFAR100 with IID and non-IID data.}
\label{skewness}
\end{figure}

We train two popular image classifers: ResNet152 \cite{b12} and VGG19 \cite{b13} on skewed data to observe the impact of unbalanced and non-IID distribution on convergence.
We induce non-IID distribution of CIFAR10 and CIFAR100 \cite{b14} by mapping a subset of labels to a unique device.
We train on 10 devices for CIFAR10 such that a single label resides on one device, while we train CIFAR100 on 25 devices by mapping 4 labels to a single device.
Using \textbf{Top-5 test accuracy} as the performance metric, Fig. \ref{skewness} shows the result of training ResNet152 on CIFAR10 and VGG19 on CIFAR100.
For comparison, we also show the corresponding performance of training with data partitioned in an IID manner.
The model quality degrades considerably on non-IID data for both models and datasets.

\subsection{Limited Memory and storage}

\begin{figure}
\subfloat[Memory util. vs. batch-size]{\includegraphics[width=0.25\textwidth]{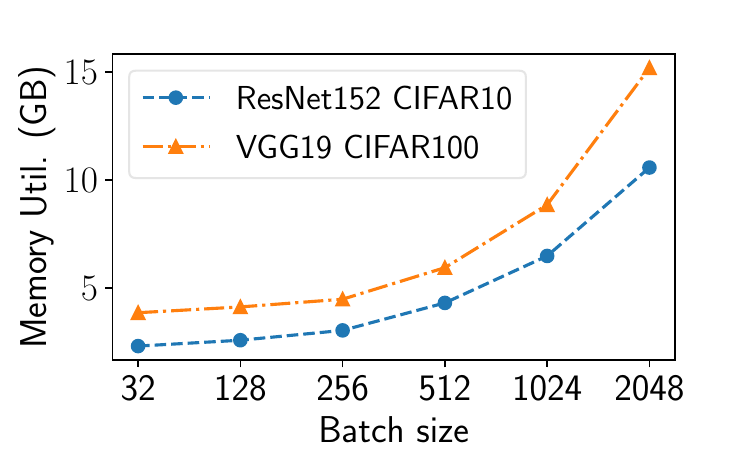}
\label{bszmem}}
\subfloat[Memory varies with SGD variant]{\includegraphics[width=0.25\textwidth]{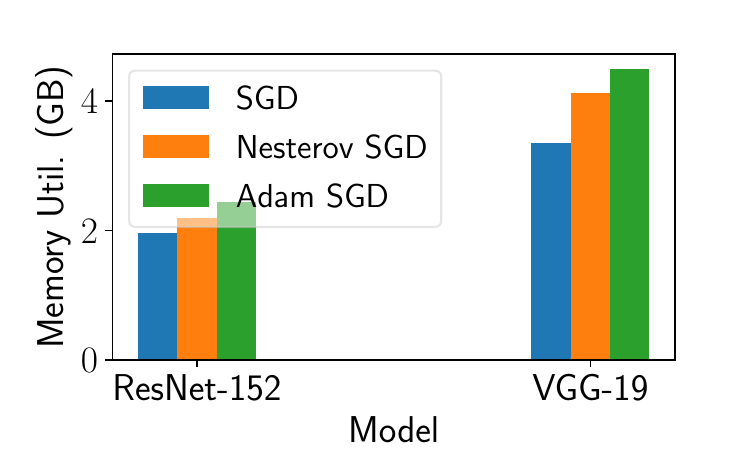}
\label{optmem}}
\caption{GPU memory utilization in DDL.} \label{memutil}
\end{figure}

With high volumes of streaming data, further processing and storage can be costly or even unfeasible due to limited physical resources.
Limited memory presents challenges even in data-center settings where GPU memory is significantly lower than system memory.
Training a neural network on a GPU requires storing model parameters (a.k.a weights), gradients computed in backward pass, activation maps as well as training batches.
Fig. \ref{memutil} shows how GPU memory utilization varies on NVIDIA V100 GPUs based on the mini-batch size and the kind of optimization used.
Keeping all other hyperparameters fixed, memory usage increases in a near-exponential fashion with batch size (Fig. \ref{bszmem}).
From Fig. \ref{optmem}, memory consumption also increases as we move from mini-batch SGD to Nesterov's momentum \cite{b15}, and then to Adam optimizer \cite{b16}.
Nesterov's momentum needs more memory than mini-batch SGD since it keeps parameter updates from the previous timestep as well.
Adam optimizer consumes even more memory since it stores both first and second order updates from previous timestep.
Even though devices designed for the edge are now more capable than ever, its still more resource constrained than dedicated data-center hardware.
This makes training neural networks on the edge even more challenging.

Since neural network training is compute-intensive, it is difficult to train models on streaming data at \emph{line-rate}.
High synchronization overhead to aggregate updates further inhibits linear scaling of DDL.
As a result, \emph{data quickly accumulates if the streaming rate is higher than the processing rate}.

The size of streaming queues can quickly blow up in distributed training, be it on-disk or in-memory.
We formulate this as follows: suppose each device $\mathit{D_{i}}$ among $[\mathit{D_{1}},\mathit{D_{2}},....\mathit{D_{n}}]$ devices has a fixed streaming rate of $\mathit{S^{(i)}}$ samples/second.
The devices collaboratively train a model with average batch-size $\mathit{b_{i}}$ (i.e., $\mathit{b_{i}} =\sum_{j=1}^{j=n}\mathit{b_{j}}/n$).
Consider a scenario where a devices' streaming rate is larger than the training batch-size, i.e., $\mathit{S^{(i)} > b_{i}}$.
A single iteration in distributed training involves calculating loss, computing gradients, aggregate and apply updates; let's denote this time on device $\mathit{i}$ as $\mathit{t_{i}}$.
At initial timestep $\mathit{ts}=0$, $\mathit{S^{(i)}}$ samples arrive each second at $\mathit{i}$ which then processes $\mathit{b_{i}}$ samples from it.
In the time $\mathit{t_{i}}$ that $\mathit{i}$ completes one training iteration, about $\mathit{(t_{i} \cdot S^{(i)})}$ more samples arrive in addition to the residual $\mathit{(S^{(i)} - b_{i})}$ that weren't used.
Thus, there are $\mathit{(S^{(i)} - b_{i})} + \mathit{t_{i}} \cdot S^{(i)}$ samples enqueued in the streaming buffer at timestep $\mathit{ts}=1$.
At timestep $\mathit{ts}=2$, there are $2\mathit{(t_{i} +1)S^{(i)}} - 2\mathit{b_{i}}$ samples in the buffer.

The queue size increases over time on account of residual samples from previous timesteps.
We generalize the number of accumulated samples $\mathit{Q_{i}}$ on device $\mathit{i}$ after $\mathit{T}$ timesteps in Eqn. (\ref{eqn:streamqueue}) and note that $\mathit{Q_{i}}$ scales linearly with $\mathit{T}$.

\begin{equation}
\mathit{Q_{i}} = \mathit{(t_{i} \cdot S_{i} - b_{i})\cdot T + S^{(i)}} \quad \forall \quad t_{i} \cdot S^{(i)} \geq b_{i}
\label{eqn:streamqueue}
\end{equation}

As a timestep corresponds to an iteration in DDL, buffer size can increase dramatically when $\mathit{T}$ is large, which is typical for neural networks to run for thousands of iterations.
To limit $\mathit{Q_{i}}$ from blowing up, one could argue to set $\mathit{b_{i}}$ to $\mathit{t_{i} \cdot S^{(i)}}$.
In that case, $\mathit{Q_{i}}$ is always equal to $\mathit{S^{(i)}}$ irrespective of the value of $\mathit{T}$ .
However, $\mathit{b_{i}}$ is a hyperparameter that may require careful tuning.
Using $\mathit{t_{i} \cdot S^{(i)}}$ as batch size can be impractical if the streaming rate is too high or too low.
A small batch-size doesn't leverage parallelism while a large $\mathit{b_{i}}$ would hurt generalization performance.

\emph{Assuming high streaming rates and considerable iteration times in DDL due to limited compute and bandwidth at the edge, Eqn. (\ref{eqn:streamqueue}) reduces to}

\begin{equation}
\mathit{Q_{i} = (T \cdot t_{i} \cdot S^{(i)} + S^{(i)}}) \quad \text{if} \quad \mathit{(t_{i} \cdot S^{(i)}}) \gg b_{i}
\label{eqn:approximatedQ}
\end{equation}

\begin{figure}
\subfloat[Streaming queue grows over time]{\includegraphics[width=0.25\textwidth]{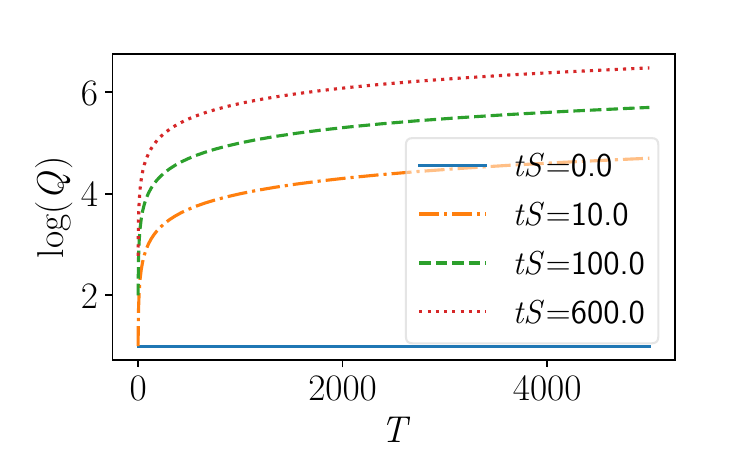}
\label{storage}}
\subfloat[Gradient synchronization time]{\includegraphics[width=0.25\textwidth]{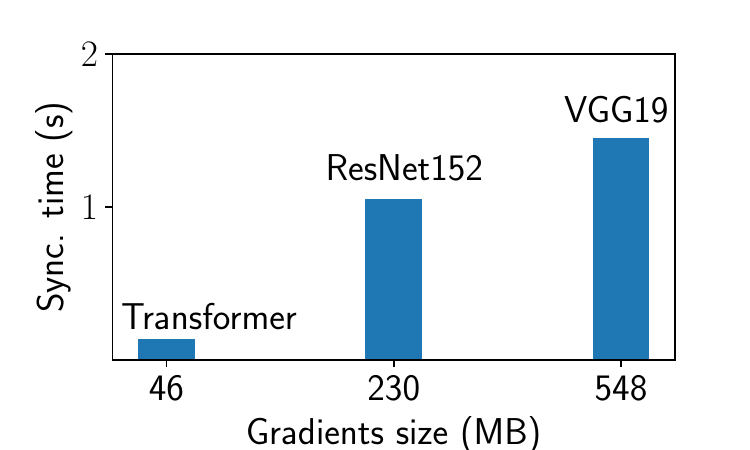}
\label{synctime}}
\caption{DDL on streaming data is limited by memory/storage as well as network bandwidth.} \label{challengesDL}
\end{figure}

We simulate how $\mathit{Q_{i}}$ increases with $\mathit{T}$ as stated in Eqn. (\ref{eqn:approximatedQ}).
The results are illustrated for different $\mathit{tS}$ values in Fig. \ref{storage}.
The y-axis takes the log (with base $10$) of the samples accumulated.
As $\mathit{tS}$ increases, so does the corresponding buffer size.
We note that when $\mathit{tS \approx 0}$, the buffer only holds $\mathit{S}$ samples at any given time.
Such a setting describes a hypothetical system where the total iteration time is negligible regardless of the streaming rate.

To gauge buffer requirements with streaming data in real-world settings, we measure the space needed to store $[32\times32]$ colored images for training ResNet152 and VGG19.
For mini-batch size $64$, the models have average iteration times of $1.2$ and $1.6$ seconds respectively.
Table \ref{storageutiltable} tabulates how training samples accumulate after $1K, 10K, 100K$ timesteps for these iteration times.
As $\mathit{T}$ increases, so does the storage requirements to hold the data.
Optimized stream processing platforms like Kafka reduce memory footprint by storing messages on-disk as partitions, and then deleting the data based on some retention policy once the messages are successfully consumed.
However, persisting data on-disk becomes unfeasible especially on the edge as data keeps accumulating with the iterations.

\begin{table}[!t]
\renewcommand{\arraystretch}{1.3}
\caption{Data accumulated with streaming in DDL}
	\centering
	\begin{tabular}{|c|c|c|c|c|c|}
	\hline
	\bfseries Model & $\mathbf{\mathit{t}}$ & $\mathit{S}$ & \multicolumn{3}{c|}{\textbf{Data accumulated at $\mathit{T}$ steps (GB)}} \\
	\cline{4-6} 
	\textbf{} & (s) & (img/s) & $\mathit{T}=10^{3}$ & $\mathit{T}=10^{4}$ & $\mathit{T}=10^{5}$ \\
	\hline
	\multirow{2}{*}{ResNet152} & \multirow{2}{*}{$1.2$} & 100 & 0.35 & 3.5 & 34.33 \\
	\cline{3-6}
	& & 600 & 2.06 & 20.6 & 200.6 \\
	\hline
	\multirow{2}{*}{VGG19} & \multirow{2}{*}{$1.6$} & 100 & 0.47 & 4.69 & 46.8 \\
	\cline{3-6}
	& & 600 & 2.75 & 27.5 & 274.83 \\
	\hline
	\end{tabular}
\label{storageutiltable}
\end{table}

\subsection{Synchronization overhead}

Although training neural networks on GPUs can significantly reduce the computation time, DDL can still incur significant overhead due to periodic gradient synchronization.
Training ResNet152 and VGG19 on 8 NVIDIA K80s takes about $80$ to $90\%$ of the total iteration time in gradient synchronization.
Additionally, communication cost tends to increase with the number of devices participating in training.
Increasing the network bandwidth brings down the synchronization cost to only a certain extent and saturating thereafter \cite{b3}.

The 8 GPUs are connected via 5Gbps ethernet for which we plot the communication time to synchronize gradients for Transformer \cite{b17}, ResNet152 and VGG19.
Fig. \ref{synctime} incurs higher communication time as the model size increases.

\section{Background and Related Work}\label{sec:bg}

\subsection{Handling unbalanced and non-IID data}

Neural network training under constraints like privacy-sensitivity, skewness due to non-IID and unbalanced data has been well studied under the premise of \emph{federated learning}.
In federated training, devices train on local, skewed data while a global shared model is learned by periodically aggregating updates from other devices.
For example, FedAvg \cite{b20} collects updates only from a fraction of total clients after certain local epochs to reduce frequent communication cost.
FedProx \cite{b24} extends FedAvg to include partial work (to address systems heterogeneity) and adds a proximal term to the local objective function to deal with statistical heterogeneity in non-IID data.
Sparse Tenary Compression (STC) \cite{b19} reduces communication by combining Terngrad \cite{b25} quantization with Top-\textit{k} \cite{b26} sparsification, and is shown to perform better than FedAvg on both IID and non-IID data.
Zhao et al. \cite{b18} account for data skewness due to weight divergence among devices' local model replica and measure it with earth mover's distance (EMD) between device-local and overall data distribution.
To facilitate development of federated learning systems and algorithms, benchmarks like FedML \cite{b21} and LEAF \cite{b22} have been developed.

\subsection{Dealing with limited memory}

As neural networks have grown in size over the years, so has their resource requirements.
The memory space to hold a model comprises of trainable parameters and a computation graph to store gradients and activation maps computed in forward-backward pass.
Micikevicius, et al. \cite{b27} proposed automatic mixed-precision (AMP) training with half-precision (16-bit) floating points to reduce the memory footprint by half.
Gradient checkpointing \cite{b28} trades memory for computation by flushing intermediate data from the computation graph to reduce memory utilization.
This comes at the cost of increased computation as flushed activation maps need to be recomputed whenever needed.
Memory consumption can be reduced by decreasing the training batch size as well.
However, this reduces parallelizability and increases the overall training time.
Deep learning frameworks like PyTorch implement \texttt{torchvision.datasets.DatasetFolder} and \texttt{ImageFolder} to avoid loading entire training data to memory by reading samples from disk one batch size at a time.   
But these dataloaders adhere to a rigid format for specifying labels in the underlying directory structure.
Thus, training a model on a particular dataset may require considerable preprocessing and designing custom dataloaders.
For massive datasets, on-disk storage can quickly blow up too as seen previously.

\subsection{Reducing communication cost}

Algorithms like FedAvg and FedProx minimize communication overhead by choosing a low-frequency, high-volume communication strategy with occasional gradient synchronization.
On the other hand, gradient compression either via sparsification, quantization or low-rank approximations uses a high-frequency, low-volume approach to reduce communication cost in DDL.
Sparsification techniques like Topk-\textit{k} \cite{b26} and Deep Gradient Compression (DGC) \cite{b29} apply sparse updates by sending only a subset of the gradients and setting remaining values to 0.
The bit-width of floating-point gradients is reduced with quantization methods.
Automatic mixed-precision (AMP) training described earlier uses half-precision gradients to achieve $2\times$ compression.
Another quantization method called Quantized SGD (QSGD) \cite{b30} quantizes gradients while balancing the trade-off between precision and accuracy.
On the other hand, Terngrad \cite{b25} limits gradients across three quantization levels [-$1$,$0$,+$1$].
Low-rank approximations like PowerSGD \cite{b36} minimize update cost by performing low-rank updates that effectively work as regularization.
All these compression techniques use a fixed compression ratio throughout training.
Using a high compression ratio incurs higher communication cost, while a small compression ratio may trim too much useful information from the gradients.
Accordian \cite{b32} dynamically compresses gradients by detecting critical regions in training by tracking gradient variance.
We extend this further by developing an adaptive compression strategy by comparing entropy loss between the original and compressed gradients.

\section{scadles}\label{sec:scadles}

We propose \emph{ScaDLES} to address the challenges described in section \ref{sec:challenges} and accelerate DDL training on heterogeneous streams in both IID and non-IID settings.

\textbf{Heterogeneous streams:} The approaches described w.r.t federated training in section \ref{sec:bg} consider either systems heterogeneity or statistical heterogeneity due to skewed and non-identical data.
Training neural networks synchronously on multiple devices with heterogeneous data streams suffers from stragglers.
A device $\mathit{i}$ among $\mathit{n}$ devices with the lowest streaming rate $\mathit{S^{(i)} \in [S^{(1)}, S^{(2)}, ...S^{(n)}]}$ can become a bottleneck depending on the mini-batch size since all other devices have to wait on $\mathit{i}$ to gather enough training samples $\mathit{b_{i}}$ and proceed an iteration.
Additionally, streaming rates can vary at intra-device level at the edge too, depending on factors like battery level, time of day, usage, etc.
This wait-time incurred due to streaming latency can thus slow down training.

To mitigate the impact of streams with lower inflows, we propose performing variable computation where we set $\mathit{b_{i} \: \forall \: i \propto S^{(i)}}$.
Thus, we minimize streaming latency by setting device batch size to its streaming rate.
As a result, some devices with high volume streams train on a large batch-size while the low volume devices use a smaller batch-size, and wait-times due to streaming latency are avoided.
Since the amount of work done on each device is different, we perform weighted aggregation rather than a simple average to get the shared global updates.
At iteration $\mathit{t}$, device $\mathit{i}$ trains with batch-size corresponding to its streaming rate $\mathit{S_{i}^{(t)}}$ and scales the computed gradients $\mathit{g_{i}^{(t)}}$ by factor $\mathit{r_{i}^{(t)}}$ and updates the parameters as:

\begin{subequations}
	\begin{equation}
		r^{(i)}_{t} = \dfrac{S^{(i)}_{t}}{\sum_{j=1}^{n}S^{(j)}_{t}} \quad : \:\: \sum_{j=1}^{n}r^{(j)}_{t} = 1.0
		\label{eqn:gradscaling}
	\end{equation}
	\begin{equation}
		\tilde{g}_{t} = \sum_{j=1}^{n} r^{(j)}_{t} \cdot g^{(j)}_{t}
		\label{eqn:gradagg}
	\end{equation}
	\begin{equation}
		w_{t+1} = w_{t} - \eta_{scaled} \cdot \tilde{g}_{t}
		\label{eqn:weightedsgd}
	\end{equation}
\label{eqn:weightedupdate}
\end{subequations}

The global batch-size with weighted gradients from Eqn. \ref{eqn:gradscaling} is $\mathit{\sum_{j=1}^{n}S^{(j)}}$.
As streaming rates can vary both inter and intra-device, so does the global batch-size.
To ensure extremely high streaming rates don't increase the global batch-size so much that it degrades generalization performance, we add a \emph{linear scaling rule} as suggested in \cite{b33, b34}.
Essentially, linear scaling adjusts the learning rate in proportion to the batch-size, i.e., learning rate is increased if the batch-size increases, and vice versa.
When the batch-size is multiplied by factor \textit{k}, multiply the base learning rate by \textit{k} as well.
If the base global batch-size is $\mathit{B}$, then we scale the base learning rate as $$\eta_{scaled}=\gamma_{scaled}\cdot \eta \;\; : \: \gamma_{scaled} = \dfrac{\sum_{j=1}^{n}S_{j}}{B}$$

Even with a linear-scaling rule for training with larger batches, model quality still may suffer with extremely large batches in high volume streams.
Likewise, using a batch-size too small is not efficient from parallelization perspective.
Thus, we set $\mathit{b_{i} = S^{(i)}}$ as long as the device batch-size is bounded in the range $\mathit{b_{min} \le b_{i} \le b_{max}}$, else we use the corresponding min-max for training. 

\textbf{Limited memory and storage:} The buffer size can grow quickly due to continuous data streams and considerable iteration times at the edge.
The accumulated data can either reside in memory like a buffered queue, and reside on-disk to reduce memory footprint like in Kafka.
By default, we could keep all the data streaming in and store it until processed successfully.
We refer to this policy as \emph{Stream Persistence}.
As seen from Table \ref{storageutiltable}, accumulated samples keep increasing over the iterations depending on the stream-rate.
Looking at Eqn. \ref{eqn:streamqueue}, buffer size grows to $\mathcal{O}(\mathit{S^{(i)}T})$ after $\mathit{T}$ iterations.
Stream persistence makes sense especially when devices have sufficient memory or storage to hold the data, like in data-centers, cloud or high-capacity fog devices.
In \emph{Stream Truncation}, we discard the residual samples and hold just enough data corresponding to the device's streaming rate $\mathit{S^{(i)}}$.
As a result, storage requirements for stream truncation is $\mathcal{O}(\mathit{S^{(i)}})$ at any given time, which is significantly smaller than stream persistence.

\textbf{Unbalanced and Non-IID data:} Fig. \ref{skewness} demonstrates how model quality degrades when training with non-identical data.
This happens since each device contains only a subset of training labels that are not representative of the entire distribution.
Thus, parameters learned by device-local model replicas are skewed, and so is the aggregated model.
To deal with non-IID data, we propose randomized \textbf{\emph{data-injection}} where a fraction of the training devices share partial training samples with other devices.
Particularly, at every iteration a device randomly chooses a subset \textbf{$\alpha$} of the total devices $\mathit{D}$ to share fraction \textbf{$\beta$} of its streaming data $\mathit{\beta S{(i) \in [\alpha D]}}$.
Together, ($\alpha, \beta$) determine what set of devices share how much of their training samples with other devices in DDL.
Data injection helps improve the overall data distribution by making the device local data more representative of the complete dataset.
However, this implies a trade-off between high model quality on account of better data distribution and privacy concern arising from moving data away from the devices.
Privacy violation is greatly minimized by choosing only a subset of devices randomly (from $\alpha$) and broadcasting only partial data (determined by $\beta$).

\textbf{High communication cost:} Federated algorithms reduce communication cost either with low-frequency, high-volume or high-frequency, low-volume communication strategy.
We focus our efforts on the latter by looking at various gradient compression techniques.
Rather than using a static compression ratio throughout training that can be detrimental to the final model accuracy, we look into adaptive compression.
Prior work keeps track of the moving average of gradient variance to detect critical regions in training \cite{b31,b32,b35}.
Gradients are large initially, but get smaller as the model evolves and training continues.
Thus, we can use low compression in the beginning and higher compression later.
We implement an adaptive compression strategy with Top-\textit{k} sparsification which compares entropy loss between compressed and uncompressed gradients.
Gradients compressed to top \textit{k\%} are used if the variance between compressed and uncompressed tensors falls below threshold $\delta$; original, uncompressed tensors are communicated otherwise.
The intuition is that if the top \textit{k}\% gradients have most of the information as the uncompressed gradients within the margin of $\delta$, then remaining gradients are relatively less meaningful that don't greatly contribute towards model update and can thus be ignored.
We track of the variance of compressed and uncompressed gradients at every iteration by keeping exponential weighted moving average (EWMA) and implement the communication rule for adaptive compression on gradients $\mathit{g}$ for a device as follows: $$\text{send}(\text{Top}\mathit{k(g)}) \;\; \text{if} \;\; \dfrac{||g|^{2} - |\text{Top}\mathit{k(g)}|^{2}|}{|g|^{2}} \le \delta \;\; \text{else} \;\; \text{send}(g)$$

\emph{Compression threshold,} denoted by $\delta$ determines the degree of relaxation we impose on the compressed tensors to be eligible for communication.
A small $\delta$ penalizes compressed tensors more severely and performs reduction only when the compressed data captures most of the relevant gradients in the original tensor.
Constraints are loose with a larger $\delta$ which allows more iterations to use compressed tensors for synchronization.

\section{Evaluation}\label{sec:eval}

\subsection{Cluster setup}

We simulate streaming data with Kafka by sampling streaming rates from uniform and normal distributions described in Table \ref{distribution_table}.
The hardware used to evaluate \emph{ScaDLES} in our experiments comprises of $4$ servers each with with $48$-core Intel Xeon E5-2650, $128$ GB system memory and $8$ NVIDIA K80 GPUs connected with $5$ Gbps ethernet.
We mimic CUDA-aware edge devices by spawning them as \texttt{nvidia-docker} containers on CentOS linux $7.9.2009$ with docker engine $20.10.17$.
Each device running as a container is allocated $4$vCPUs, $12$ GB system memory and $1$ K80 GPU running NVIDIA driver $465.19.01$ on CUDA $11.3$ and PyTorch $1.10.1$.
We create a docker swarm network on the $5$ Gbps network interface to facilitate communication for gradient synchronization among containers.

\subsection{Data, models and hyperparameters}

We evaluate two popular neural networks across different streaming distributions, training dataset and cluster configurations.
ResNet152 uses SGD optimizer with momentum $0.9$ and weight decay $0.0001$ while adopting a learning rate schedule with initial lr $0.1$ that decays by $0.2$ after $75$, $150$ and $225$ epochs.
We also train VGG19 momentum SGD of $0.9$ and weight decay $0.0005$ with an intial lr $0.01$ that decays by $0.3$ after $75$, $150$ and $200$ epochs.
The model quality for both neural networks is measured by the \emph{Top-5 test accuracy}.

The cluster setup for both IID and non-IID data is outlined in Table \ref{modeldescription}.
Training with IID data is performed on $16$ devices where each device is equipped with a K80 GPU.
We partition non-IID data by mapping a device to a unique subset of labels.
Non-IID CIFAR10 is trained on $10$ devices where each device contains only a single label.
We train non-IID CIFAR100 on $25$ devices such that each device is mapped to $4$ unique labels.

\begin{table}[!t]
\renewcommand{\arraystretch}{1.3}
\caption{Neural networks evaluated}
	\centering
	\begin{tabular}{|c|c|c|c|c|}
	\hline
	\bfseries Model & \bfseries Parameters & \bfseries Data & \bfseries Devices & \bfseries Labels/device\\
	\hline
	\multirow{2}{*}{ResNet152} & \multirow{2}{*}{$60.2M$} &  IID Cifar10 & 16 & 10 \\
	\cline{3-5}
	& & nonIID Cifar10 & 10 & 1 \\
	\hline
	\multirow{2}{*}{VGG19} & \multirow{2}{*}{$143.7M$} & IID Cifar100 & 16 & 100 \\
	\cline{3-5}
	& & nonIID Cifar100 & 25 & 4 \\
	\hline
	\end{tabular}
\label{modeldescription}
\end{table}

\subsection{Streaming data for DDL}

We use Apache Kafka v$3.1.0$ to spawn a docker container that runs a broker as well as producers.
The broker-producer container is allocated $16$vCPUs, $32$ GB system memory and \emph{no} GPU since it doesn't participate in model training.
We configure the container with $8$ network threads, $4$ IO-threads and $1$ partition per topic.
The sole purpose of this container is to host the Kafka broker and launch multiple producer processes such that each process publishes to a unique topic corresponding to a device.
Thus, there are as many topics as the number of training devices.
Each producer process controls the streaming rate corresponding to a device's topic.
As for data consumption, the training devices have a kafka consumer running on them.
The consumer implements a custom PyTorch dataloader that batches the data and integrates into a typical training loop that is common in deep learning training.

\begin{figure}
\subfloat[100 samples/sec]{\includegraphics[width=0.25\textwidth]{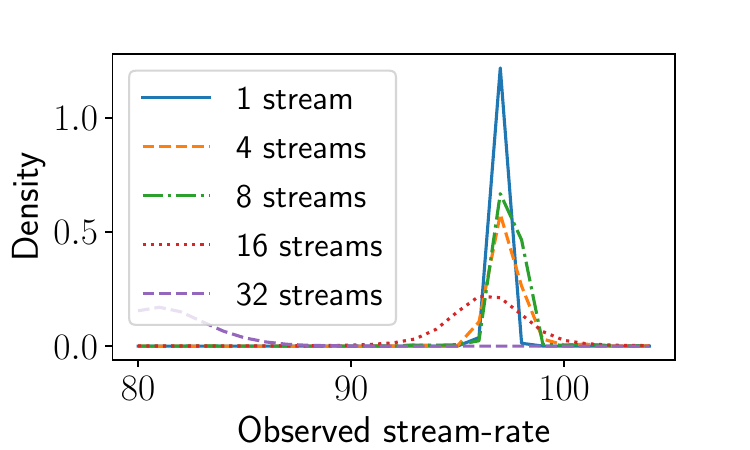}
\label{stream100events}}
\subfloat[600 samples/sec]{\includegraphics[width=0.25\textwidth]{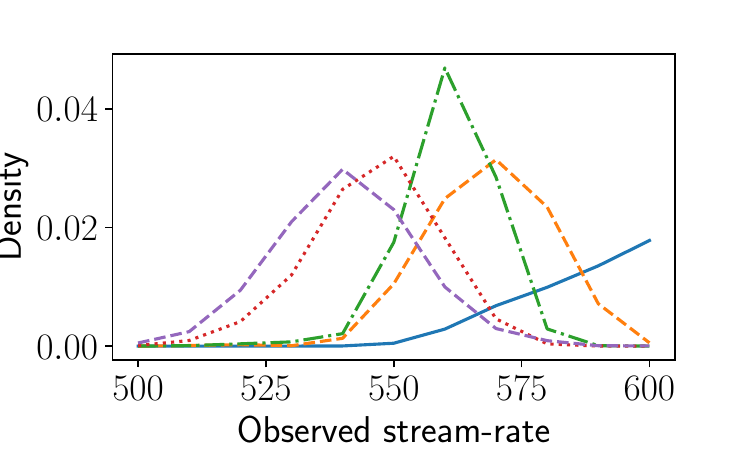}
\label{stream600events}}
\caption{Effective streaming rates achieved when scaling to multiple topics.} \label{fig:streamsimulate}
\end{figure}

Since we run multiple producers from a single container (and not a separate container for every producer), we measure the effective streaming rate achieved with our proposed setup to ensure we meet the target stream-rates in our experiments.
We scale up the number of concurrent producers (i.e., topics) and measure the observed streaming rate.
For e.g., $32$ streams in Fig. \ref{stream100events} imply $32$ producers publishing to $32$ topics at $100$ samples/sec.
Fig. \ref{fig:streamsimulate} shows the density estimates of observed streaming rates with targets of $100$ and $600$ samples/sec.
For each target rate, we scale up the number of concurrent producers to $1,4,8,16$ and $32$.
We achieve nearly the same target of $100$ samples/sec as shown in Fig. \ref{stream100events}.
For the $600$ samples/sec target, the effective streaming rate decreases noticeably beyond $16$ concurrent streams.
We could likely improve this by increasing the number of network threads and partitions per topic, but this setup sufficed for the evaluations we perform in this paper.

\subsection{Weighted aggregation in heterogeneous streams}

We use a batch-size corresponding to a devices' streaming rate in \emph{ScaDLES} to avoid wait-times on account of possible streaming latency.
To enforce bounds on the batch-size used, we set $\mathit{b_{min}}$ and $\mathit{b_{max}}$ to $8$ and $1024$ respectively, although stream-rate for any device remains within this range regardless of the streaming distribution.
The waiting time can especially be long in highly heterogeneous streams when a devices' streaming rate is lower than the mini-batch size configured prior training.
There is no waiting time high-volume streams with low-batch size settings.
However, the buffer size can grow quickly over time in that case.
We compare \emph{ScaDLES} with conventional DDL training for batch-size $64$ irrespective of the device streaming rates.
We look at the convergence curves and buffer buildup over training epochs to compare the two.
The streaming rates for the 16 devices were sampled from the distributions outlined in section \ref{sec:challenges}.
Uniform distributions are more heterogeneous compared to normal distributions ($2/3$rd values lie within $1$ standard deviation from the mean in the latter).

\begin{figure}
\subfloat[$\mathit{S_{1}}$ distribution]{\includegraphics[width=0.25\textwidth]{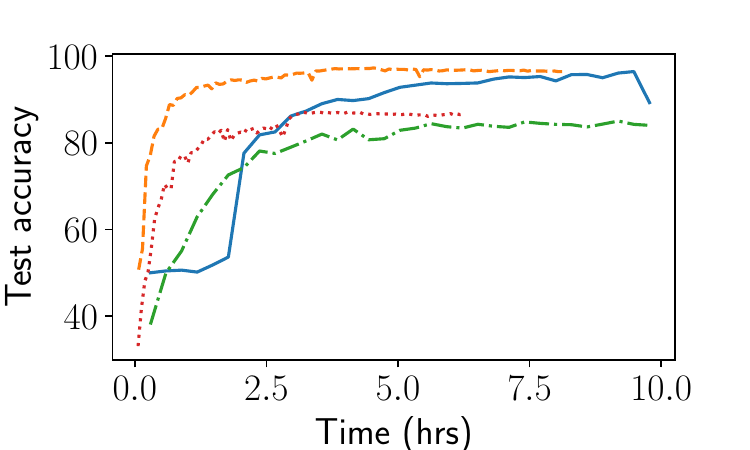}
\label{uniform_s1}}
\subfloat[$\mathit{S_{2}}$ distribution]{\includegraphics[width=0.25\textwidth]{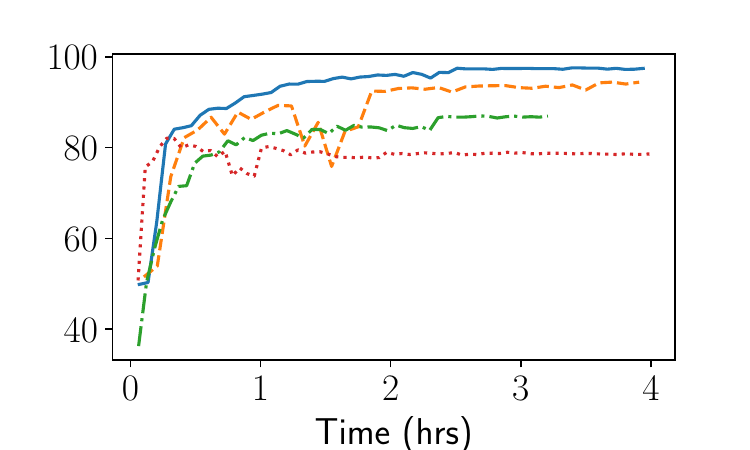}
\label{uniform_s2}}
\hspace{0.01cm}
\subfloat[$\mathit{S_{1}'}$ distribution]{\includegraphics[width=0.25\textwidth]{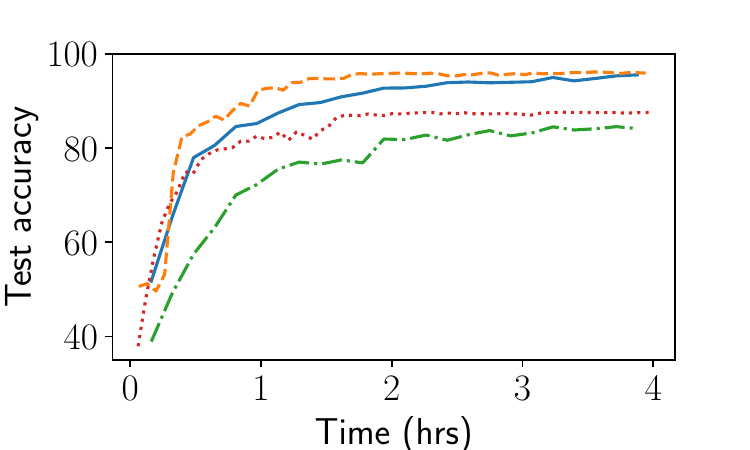}
\label{uniform_s1dash}}
\subfloat[$\mathit{S_{2}'}$ distribution]{\includegraphics[width=0.25\textwidth]{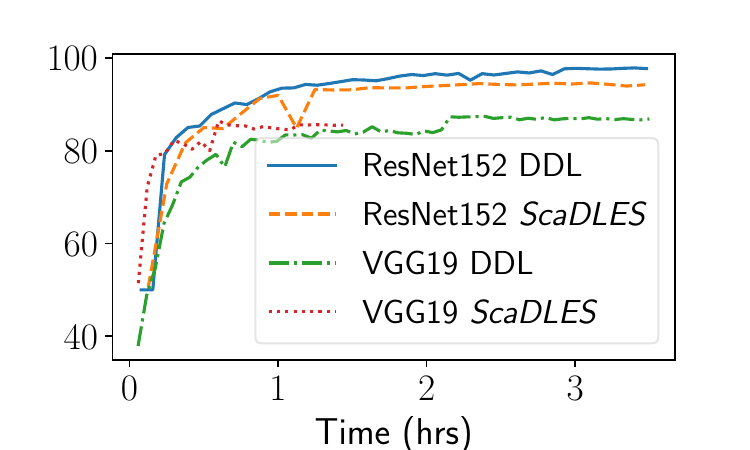}
\label{uniform_s2dash}}
\caption{Convergence in conventional DDL vs. weighted aggregation approach in \emph{ScaDLES}.}\label{fig:het_streams}
\end{figure}

Fig. \ref{uniform_s1} shows results for \emph{ScaDLES} and conventional DDL using device stream-rates sampled from $\mathit{S_{1}}$ that converges $3.33\times$ and $1.92\times$ faster in \emph{ScaDLES}.
Conventional DDL converged with higher final accuracy in $\mathit{S_{2}}$ due to large batches used for training by \emph{ScaDLES} with this distribution; about $4.5$K in \emph{ScaDLES} compared to only $1$K in DDL.
Linearly scaling the learning rate at batches this large did not significantly improve \emph{ScaDLES}' generalization performance.
Devices sampled from $\mathit{S_{1}'}$ achieved around $3.6\times$ and $4\times$ speedup with \emph{ScaDLES} while still achieving higher final accuracy.
Using larger batches and linear scaling proved to be beneficial in this case.
Lastly, Fig. \ref{uniform_s2dash} uses $\mathit{S_{2}'}$ distribution where ResNet152 performs similarly for both \emph{ScaDLES} and DDL, while VGG19 performs better with our approach.

\subsection{Managing limited memory and storage}

\begin{figure}
\subfloat[$\mathit{S_{1}}$ distribution]{\includegraphics[width=0.25\textwidth]{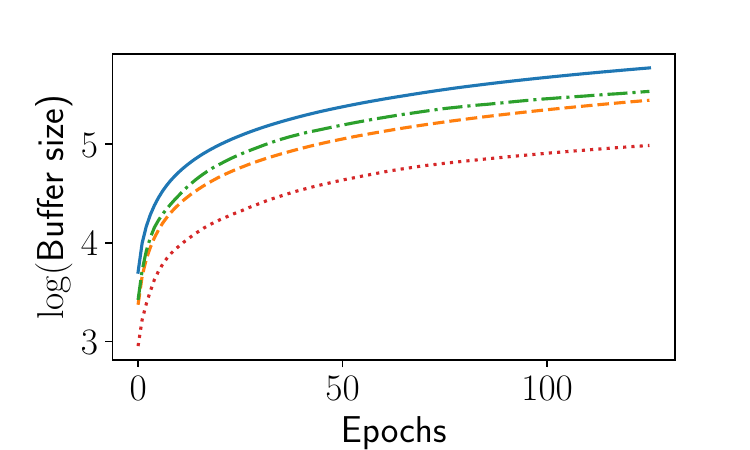}
\label{uniform_s1mem}}
\subfloat[$\mathit{S_{2}}$ distribution]{\includegraphics[width=0.25\textwidth]{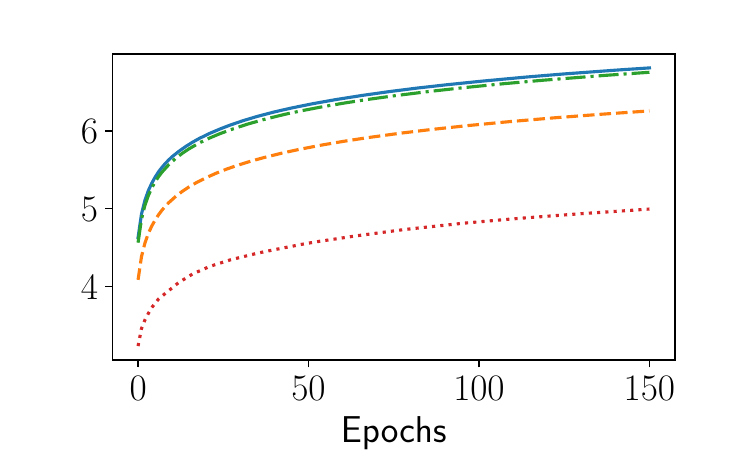}
\label{uniform_s2mem}}
\hspace{0.01cm}
\subfloat[$\mathit{S_{1}'}$ distribution]{\includegraphics[width=0.25\textwidth]{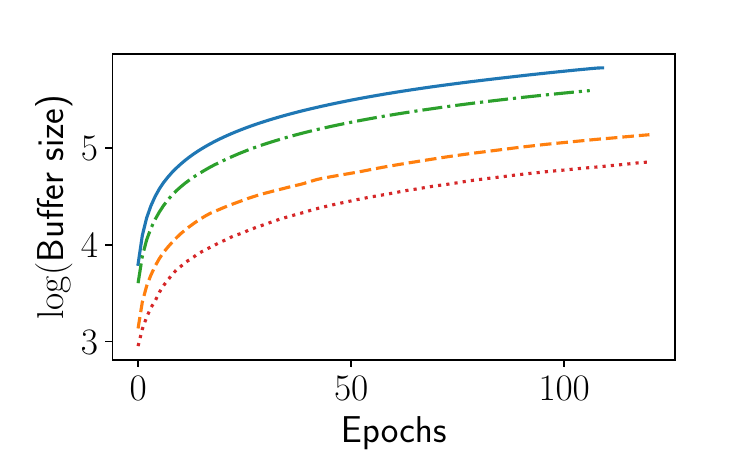}
\label{uniform_s1dashmem}}
\subfloat[$\mathit{S_{2}'}$ distribution]{\includegraphics[width=0.25\textwidth]{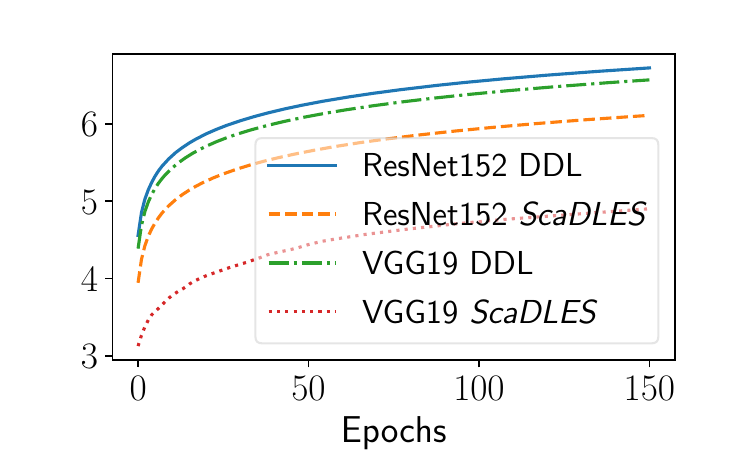}
\label{uniform_s2dashmem}}
\caption{Buffer size increases  with training iterations}\label{fig:memutil_streams}
\end{figure}

We first look at how streaming data gets accumulated in a device with the default persistence policy.
For the same runs described in the previous section, we plot how samples get accumulated over the iterations for different sampling distributions in Fig. \ref{fig:memutil_streams}.
We plot logarithm of the accumulated samples with base $10$.
The buffer size is smaller in \emph{ScaDLES} compared to DDL training for the same persistence policy.
This is because \emph{ScaDLES} uses batch-size $\mathit{S^{(i)}}$ while we use a smaller batch-size $64$ in conventional DDL.
$\mathit{S_{2}}$ and $\mathit{S_{2}'}$ have larger buffer sizes since they represent higher volume streams compared to $\mathit{S_{1}}$ and $\mathit{S_{1}'}$.
DDL occupies $2\times$ and $3.5\times$ more space with ResNet152 and VGG19 in $\mathit{S_{1}}$.
\emph{ScaDLES} holds $3.6\times$ and $641\times$ less data than DDL for $\mathit{S_{2}}$.
Comparing Fig. \ref{uniform_s2} and Fig. \ref{uniform_s2mem}, we see the lower buffer size in \emph{ScaDLES} came at the cost of lower final accuracy due to large-batch training.
\emph{ScaDLES} has $4.7\times$ and $5\times$ smaller buffer in $\mathit{S_{1}'}$, and $3.9\times$ and $42\times$ lesser data with $\mathit{S_{2}'}$ distribution.

Although \emph{ScaDLES} accumulates lesser samples than conventional DDL, we can further lower the buffer size for continuous data streams.
With stream truncation, data in buffer exceeding the samples that just streamed in is simply discarded.
The buffer size with truncation policy is constant as long as the streaming rate is continuous.
On the other hand, persistence policy grows with each passing iteration.
Table \ref{table:persist_trun} shows the final buffer size to reach $95$\% and $84$\% accuracy on ResNet152 and VGG19.
The table also reports reduction with truncation relative to persistence policy.
We observed buffer reductions from $850\times$ to $9400\times$ depending on the distribution.

\begin{table}[!t]
\renewcommand{\arraystretch}{1.3}
\caption{Buffer-size reduction with truncation policy}
	\centering
	\begin{tabular}{|c|c|c|c|c|}
	\hline
	\bfseries Dist. & \bfseries Model & \bfseries Persistence & \bfseries Truncation & \bfseries Reduction \\
	\hline
	\multirow{2}{*}{$\mathit{S_{1}}$} & ResNet152 & $2.9 \times 10^{5}$ & $129$ & $2238\times$ \\
	\cline{2-5}
	& VGG19 & $1 \times 10^{5}$ & $118$ & $848\times$ \\
	\hline
	\multirow{2}{*}{$\mathit{S_{2}}$} & ResNet152 & $4.36 \times 10^{6}$ & $633$ & $6889\times$ \\
	\cline{2-5}
	& VGG19 & $4 \times 10^{6}$ & $523$ & $7830\times$ \\
	\hline
	\multirow{2}{*}{$\mathit{S_{1}'}$} & ResNet152 & $6.2 \times 10^{5}$ & $143$ & $4340\times$ \\
	\cline{2-5}
	& VGG19 & $3.7 \times 10^{5}$ & $129$ & $2861\times$ \\
	\hline
	\multirow{2}{*}{$\mathit{S_{2}'}$} & ResNet152 & $3.6 \times 10^{6}$ & $384$ & $9429\times$ \\
	\cline{2-5}
	& VGG19 & $2.5 \times 10^{6}$ & $360$ & $6956\times$ \\
	\hline
	\end{tabular}
\label{table:persist_trun}
\end{table}

\subsection{Data-injection for non-IID and skewed data}

\begin{figure}
\subfloat[ResNet152]{\includegraphics[width=0.25\textwidth]{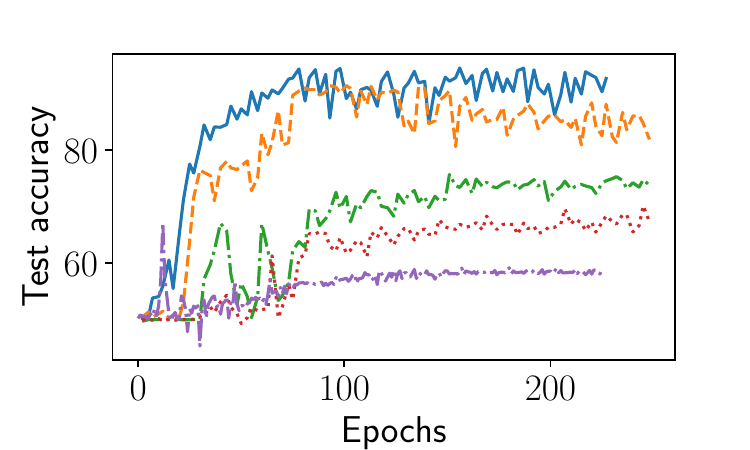}
\label{skew_resnet}}
\subfloat[VGG19]{\includegraphics[width=0.25\textwidth]{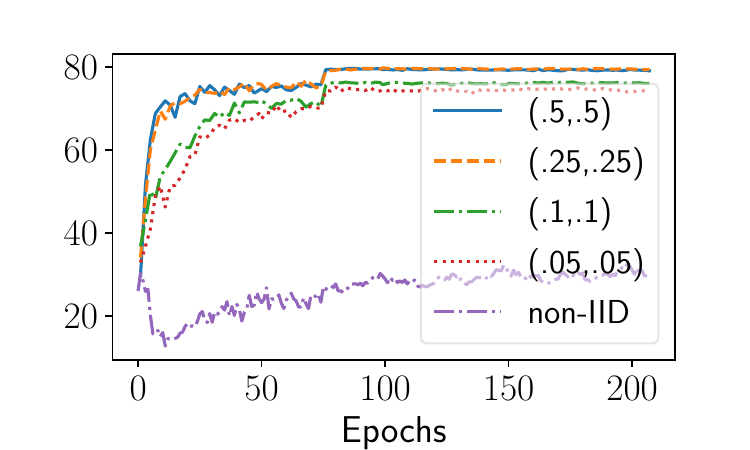}
\label{skew_vgg}}
\caption{Different ($\alpha,\beta$) configurations for data-injection in non-IID training.}\label{fig:skewness_pq}
\end{figure}

Data-injection helps improve overall model quality when dealing with non-IID data.
The degree of data-injection is determined by ($\alpha, \beta$) parameters that determine the subset of devices to send partial data.
We evaluate four ($\alpha, \beta$) sets in \emph{ScaDLES}: $(0.5,0.5)$, $(0.25,0.25)$, $(0.1,0.1)$ and $(0.05,0.05)$.
A value of $(0.5,0.5)$ means half of the devices share half of the samples in their current batch.
We plot the convergence curves for different streaming distributions in Fig. \ref{fig:skewness_pq} and note significantly better performance than training merely with non-IID data.

\begin{figure}
\subfloat[$\mathit{S_{1}}$ distribution]{\includegraphics[width=0.25\textwidth]{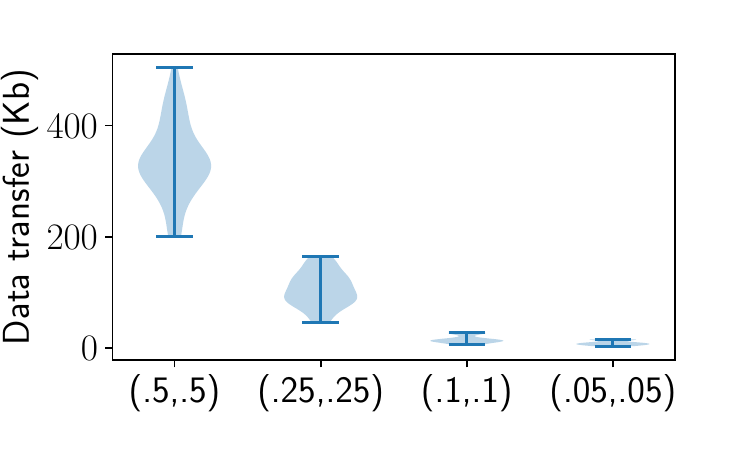}
\label{injectoverhead_s1}}
\subfloat[$\mathit{S_{2}}$ distribution]{\includegraphics[width=0.25\textwidth]{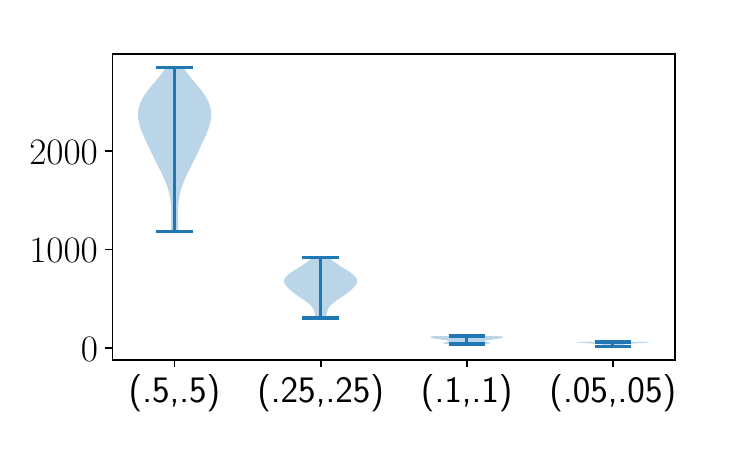}
\label{injectoverhead_s2}}
\hspace{0.01cm}
\subfloat[$\mathit{S_{1}'}$ distribution]{\includegraphics[width=0.25\textwidth]{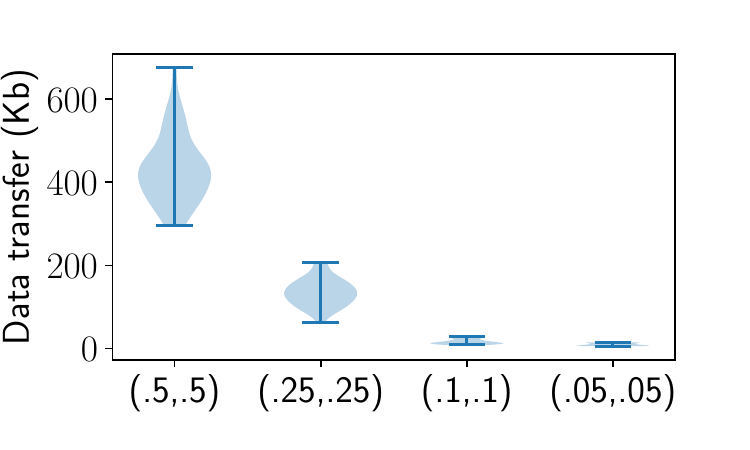}
\label{injectoverhead_s1dash}}
\subfloat[$\mathit{S_{2}'}$ distribution]{\includegraphics[width=0.25\textwidth]{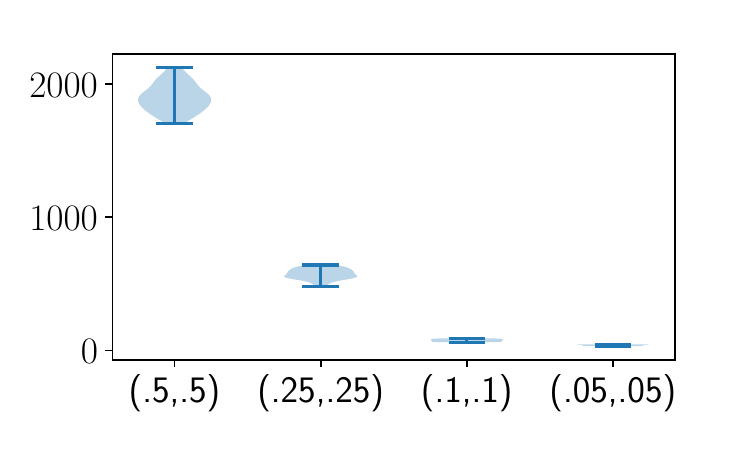}
\label{injectoverhead_s2dash}}
\caption{Data transfer overhead at each iteration to handle non-IID data with data-injection.}\label{fig:injection_overhead}
\end{figure}

Some additional networking cost is associated with data-injection as a subset of devices send partial data to other devices.
For CIFAR10 and CIFAR100 datasets, each sample is an image $3$ Kilobytes in size.
For different streaming distributions and ($\alpha,\beta$) parameters, we look at data exchange among the devices in Fig. \ref{fig:injection_overhead}.
The overhead is minimal and ranges anywhere from $150$ to $2000$ kilobytes on average for each training iteration.

\subsection{Adaptive compression}

We look at reduction in the overall communication volume, i.e., cumulative single-precision floats communicated achieving target accuracy to evaluate the performance of adaptive compression in \emph{ScaDLES}.
\textbf{Compression ratio (CR)} measures the degree of compression by comparing the tensor size of compressed gradients to that of the original tensors.
CR of $0.1$ means compressed tensors are $1/10$-th the size of uncompressed gradients.
The update size of ResNet152 ($230$ MB) and VGG19 ($548$ MB) at this CR reduces to just $23$ and $55$ MB respectively. 

Using the adaptive compression rule described in section \ref{sec:scadles}, we measure the usage of compressed gradients for communication with \textbf{Compression-to-No-Compression (CNC) ratio}.
CNC ratio compares iterations using compressed gradients for communication to the total iterations used throughout training.
The latter includes the iterations that use compression as well iterations that use the original, uncompressed gradients for communication: $$\text{CNC ratio} = \frac{T_{compressed}}{T_{compressed} + T_{uncompressed}}$$

CNC ratio of $0$ means that compressed tensors \emph{were not} used even for a single iteration, while CNC of $1.0$ implies all training iterations used \emph{only} the compressed tensors for exchange since there are no iterations that used the original gradients for communication.
To measure the impact of adaptive compression on model convergence, we tabulate the CNC ratio, accuracy and overall reduction in communication volume for different (CR, $\delta$) configurations in Table \ref{table:adaptive_comp}.
ResNet152 running with CR $0.1$ and any $\delta$ beyond $0.2$ is faster as it converges by exchanging fewer floats.
A $\delta$ of $0.1$ barely used any compression for the two CRs in either of the models.
The most communication efficient configuration for ResNet152 used CR $0.01$ and $\delta$ $0.4$, although it results in slightly lower test accuracy.
The pattern of low communication overhead accompanied with degradation in final accuracy was also observed in VGG19 for CR $0.1$ and $\delta$ $0.4$.
The CNC ratio is high in both cases implying compression is enabled for most iterations and thus, accuracy drop can be attributed to model degradation commonly associated with compression.
An interesting observation in VGG19 using CR $0.01$ is the communication volume is same across all $\delta$ values and the total floats exchanged is the same as training without any compression.
This means the adaptive strategy is not using compression in this configuration, which is further corroborated by the CNC ratio that is $0$ across all $\delta$.

\begin{table}[!t]
\renewcommand{\arraystretch}{1.3}
\caption{communication reduction in adaptive compression}
	\centering
	\begin{tabular}{|c|c|c|c|c|c|}
	\hline
	\bfseries Model & \bfseries CR & \bfseries $\delta$ & \bfseries CNC ratio & \bfseries Accuracy & \bfseries Floats sent \\
	\hline
	\multirow{8}{*}{ResNet152} & 	\multirow{4}{*}{$0.1$} & $0.1$ & $0.29$ & 97.55\% & $4.43 \times 10^{11}$ \\
	\cline{3-6}
	& & $0.2$ & $0.99$ & 96.81\% & $0.56 \times 10^{11}$ \\
	\cline{3-6}
	& & $0.3$ & $1.0$ & 98.41\% & $0.4 \times 10^{11}$ \\
	\cline{3-6}
	& & $0.4$ & $1.0$ & 98.57\% & $0.4 \times 10^{11}$ \\
	\cline{2-6}
	\cline{2-6}
	& 	\multirow{4}{*}{$0.01$} & $0.1$ & $0$ & 97.39\% & $6.02 \times 10^{11}$ \\
	\cline{3-6}
	& & $0.2$ & $0.17$ & 97.47\% & $4.99 \times 10^{11}$ \\
	\cline{3-6}
	& & $0.3$ & $0.43$ & 96.72\% & $2.56 \times 10^{11}$ \\
	\cline{3-6}
	& & $0.4$ & $0.99$ & 94.97\% & $6.32 \times 10^{8}$ \\
	\cline{2-6}
	\hline
	\multirow{8}{*}{VGG19} & \multirow{4}{*}{$0.1$} & $0.1$ & $0$ & 85.45\% & $1.3 \times 10^{12}$ \\
	\cline{3-6}
	& & $0.2$ & $0.08$ & 84.74\% & $1.19 \times 10^{12}$ \\
	\cline{3-6}
	& & $0.3$ & $1.0$ & 81.91\% & $1.3 \times 10^{10}$ \\
	\cline{3-6}
	& & $0.04$ & $1.0$ & 81.78\% & $1.3 \times 10^{10}$ \\
	\cline{2-6}
	\cline{2-6}
	& 	\multirow{4}{*}{$0.01$} & $0.1$ & $0$ & 84.68\% & $1.3 \times 10^{12}$ \\
	\cline{3-6}
	& & $0.2$ & $0$ & 83.98\% & $1.3 \times 10^{12}$ \\
	\cline{3-6}
	& & $0.3$ & $0$ & 83.94\% & $1.3 \times 10^{12}$ \\
	\cline{3-6}
	& & $0.4$ & $0.004$ & 84.39\% & $1.29 \times 10^{12}$ \\
	\hline
	\end{tabular}
\label{table:adaptive_comp}
\end{table}

\subsection{Overall performance of ScaDLES}

Last, we look at the overall performance gains in \emph{ScaDLES} by combining weighted aggregation in heterogeneous streams, data-injection for non-IID data, buffer reduction with stream truncation and reducing communication with adaptive compression (using CR $0.1$ and $\delta$ of $0.3$ in our final evaluation).
We compare against conventional DDL with fixed batch-size $64$, persistence policy and the same training schedule as \emph{ScaDLES} described in section \ref{sec:eval}.

For the same streaming distribution, \emph{ScaDLES'} performance is measured relative to conventional DDL in terms of drop in test accuracy, reduction in buffer size using truncation policy (in Gigabytes) and overall training speedup to convergence w.r.t wall-clock time.
A negative accuracy drop means \emph{ScaDLES} achieved lower accuracy by that margin.
Table \ref{table:overall_speedup} shows the results for IID training.
ResNet152 trains on \emph{ScaDLES} with a maximum of drop of $0.32\%$ in final model accuracy compared to conventional DDL with stream-rates sampled from $\mathit{S_{2}}$.
Training with \emph{ScaDLES} is also much faster; ranging from $1.15\times$ to $3.29\times$ faster than DDL.
For high-volume streams $\mathit{S_{2}}$ and $\mathit{S_{2}'}$,  truncation policy saves up to $5.9$ GB in the occupied buffer-size.
\emph{ScaDLES} achieved lower final accuracy in VGG19 by as much as $4.18\%$.
We observed that VGG19 is more sensitive to the combination of \emph{ScaDLES}' large-batch training and adaptive compression than ResNet152.
However, VGG19 still reduces buffer-size by up to $3.91$ GB and reduces wall-clock training time by $1.56\times$ to $2.83\times$ over conventional DDL.

\begin{table}[!t]
\renewcommand{\arraystretch}{1.3}
\caption{scadles' performance gains over conventional ddl}
	\centering
	\begin{tabular}{|c|c|c|c|c|}
	\hline
	\bfseries Model & \bfseries Dist. & \bfseries Acc. drop & \bfseries Buffer red. (GB) & \bfseries Speedup \\
	\hline
	\multirow{4}{*}{ResNet152} & $\mathit{S_{1}}$ & $-0.06$\% & $0.6$ & $\mathbf{1.89\times}$ \\
	\cline{2-5}
	& $\mathit{S_{2}}$ &   $-0.32$\% & $5.9$ & $\mathbf{1.15\times}$ \\
	\cline{2-5}
	& $\mathit{S_{1}'}$ &   $-0.13$\% & $0.8$ & $\mathbf{3.29\times}$ \\
	\cline{2-5}
	& $\mathit{S_{2}'}$ &   $-0.21$\% & $4.03$ & $\mathbf{1.42\times}$ \\
	\hline
	\multirow{4}{*}{VGG19} & $\mathit{S_{1}}$ & $-1.93$\% & $0.26$ & $\mathbf{1.56\times}$ \\
	\cline{2-5}
	& $\mathit{S_{2}}$ &   $-4.18$\% & $3.91$ & $\mathbf{2.83\times}$ \\
	\cline{2-5}
	& $\mathit{S_{1}'}$ &   $-2.03$\% & $0.35$ & $\mathbf{2.06\times}$ \\
	\cline{2-5}
	& $\mathit{S_{2}'}$ &   $-1.59$\% & $2.58$ & $\mathbf{2.13\times}$ \\
	\hline
	\end{tabular}
\label{table:overall_speedup}
\end{table}

As for training with non-IID data, conventional DDL was unable to reach the same convergence targets as \emph{ScaDLES}' data-injection strategy for either of the neural networks.
Conventional DDL with non-IID data saturated ResNet152 at $56\%$ test accuracy, while VGG19 did not improve beyond $35\%$.
For the same non-IID training, \emph{ScaDLES} achieved at least $93.6\%$ and $77.8\%$ accuracy for ResNet152 and VGG19 across the four stream-rate distributions.

\section{conclusion}

This paper presents the notion of training neural networks efficiently over streaming data at the edge.
Streaming data presents challenges affecting both parallel and statistical efficiency of distributed training.
\emph{ScaDLES} addresses the problem of heterogeneous streaming rate among devices with weighted aggregation where each device trains on the samples accumulated in the stream and avoids wait-time or large buffer accumulation.
Since it is difficult to perform training at line-rate, samples streaming into the device can accumulate over time.
\emph{ScaDLES} uses a simplistic truncation policy to keep the buffer size in check.
Devices on the edge commonly have non-IID and unbalanced data.
Data-injection strategy improves model convergence significantly in such scenarios.
Lastly, we propose an adaptive compression technique to deal with limited bandwidth on the edge and high communication cost in large deep learning models.
We simulate different degrees of streaming heterogeneity by sampling from both uniform and normal distributions and evaluate popular image classifiers.
Our empirical evaluation shows that \emph{ScaDLES} can converge anywhere from $1.15\times$ to $3.29\times$ faster than DDL.
At the same time, \emph{ScaDLES} reduces the number of accumulated samples in the buffer by $848\times$ to $9429\times$.


\end{document}